\documentclass[twocolumn,prl,showpacs,preprintnumbers,amsmath,amssymb,showkeys]{revtex4}
\usepackage{graphicx}
\usepackage{dcolumn}
\usepackage{bm}

\begin{document}

\preprint{PRL, 2006}

\title{Dissolution arrest and stability of armored bubbles}

\author{Manouk Abkarian$^{1}$}
\altaffiliation[Present address: ]{Laboratoire des Collo\"{i}des,
Verres et Nanomat\'eriaux, UMR5587, CC26, UMII, 34095 Montpellier Cedex 5, France}
\email{abkarian@lcvn.univ-montp2.fr}
\author{Anand Bala Subramaniam$^{1}$}
\author{Shin-Hyun Kim$^{2}$}
\author{Ryan Larsen$^{1}$}
\author{Seung-Man Yang$^{2}$}
\author{Howard A. Stone$^{1}$}
\email{has@deas.harvard.edu} \affiliation{\rm 1. Division of
Engineering and Applied Sciences, Harvard University, Pierce Hall,
29 Oxford Street, Cambridge, Massachusetts 02138, USA}
\affiliation{\rm 2. Center for Integrated Optofluidic Systems and
Department of Chemical and Biomolecular Engineering, Korea Advanced
Institute of Science and Technology, Daejeon, 305-701, Korea}

\begin{abstract}

Dissolving armored bubbles stabilize with nonspherical shapes by
jamming the initially Brownian particles adsorbed on their
interfaces. In a gas-saturated solution, these shapes are
characterized by planar facets or folds for decreasing ratios of the
particle to bubble radii. We perform numerical simulations that
mimic dissolution, and show that the faceted shape represents a
local minimum of energy during volume reduction. This minimum is
marked by the vanishing of the Laplace overpressure $\Delta P$,
which together with the existence of a $V$-interval where $d\Delta
P/dV>0$ guarantees stability against dissolution. The reduction of
$\Delta P$ is due to the saddle-shape deformation of most of the
interface which accompanies the reduction in the mean curvature of
the interface.

\end{abstract}

\pacs{Valid PACS appear here} \keywords{Armored bubbles, Jamming,
Foams, Facets}

\maketitle

It is well established that colloidal particles adsorbed on bubble
surfaces can increase bubble
\cite{Ramsden,Johnson81,Du03,Dickinson04} and foam
\cite{Alargova04,Binks05} lifetimes by several orders of magnitude
in gas-saturated solutions. This significant increase in stability
has potential applications in fields as diverse as biomedicine
\cite{Schutt05}, materials science \cite{Krachelsky01}, mineral
flotation \cite{Adamson97} and food processing \cite{Gibbs99}.
Nevertheless, in spite of the many reports of long-lived foams and
bubbles covered with particles (armored bubbles), the mechanism of
armored bubble stabilization remains an open question.

In this Letter, we seek to address the issue of stabilization using
both experimental and numerical approaches. We begin by considering
the dissolution of a single component gas bubble in a liquid
saturated with the same gas. The driving force for dissolution is
the pressure difference created inside the bubble due to the mean
curvature, $H$ and the surface tension $\gamma$ that exists at the
bubble surface. This Laplace pressure difference, $\Delta$$P=2\gamma
H$, is positive for bubbles, and thus gas in the bubble has a higher
chemical potential than the gas dissolved in the liquid. On
thermodynamic grounds, dissolution in saturated solutions can be
slowed down if this overpressure is reduced or even stopped if the
overpressure is eliminated. Indeed, the modest increase in bubble
lifetimes for surfactant-coated interfaces is due to the lowering of
the gas-liquid surface tension, with more significant increases in
lifetime occurring for bubbles covered with gelled monolayers of
lipids \cite{Needham}. However, unlike molecular surfactants or
lipids, colloidal particles are not amphiphillic, and thus do not
change the surface tension. This then raises the question of how do
the adsorbed particles reduce the overpressure of the bubble?

Several related studies provide some insight. Numerical studies of
fluid infiltration of granular media have shown a concave
deformation of the infiltrating interface as a function of the
volume and contact angle of the particles \cite{Hilden03}. A 2D
analytical study of armored bubbles found that the ``particles" pack
into a circular shape, while the interface becomes flat \cite{Kam99}
and such a flat interface is stable to perturbations
\cite{Subramanian05}. As we show below, armored bubbles stabilize in
various non-spherical and irregular shapes, whose stability can be
understood in terms of the interface shapes, characterized by the
mean and Gaussian curvatures at the scale of individual particles.

We perform our experiments with negatively charged, surfactant-free
fluorescent latex particles (Interfacial Dynamics). Partially coated
bubbles were produced as described in \cite{Subramaniam06}. An
aqueous sample containing the bubbles was placed on a microscope
slide and viewed with an inverted microscope. The small size of the
sample ensures that it is saturated with gas. All experiments were
carried out at room temperature. The images were acquired with a CCD
camera and treated with Image J to obtain a projection of the
visible surface of the armored bubble (for details see
\cite{supplement}).

\begin{figure}
\includegraphics[width=8.5 cm]{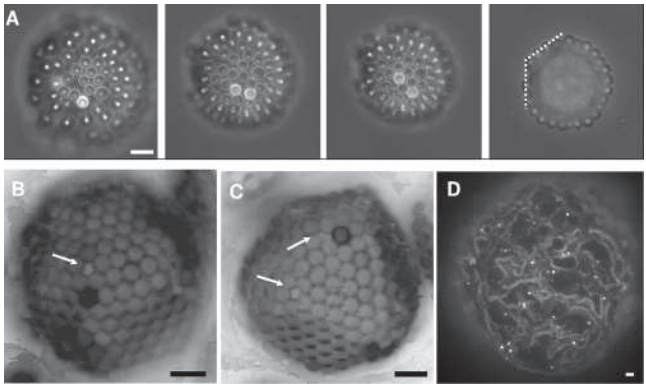}
\caption{\label{fig1}(A) Dissolution of a partially-covered bubble;
3 s between each frame. Interparticle distances are reduced and the
bubble develops planar facets as it stabilizes (white dashed lines).
(B-D) Various stable faceted and crumpled shapes of armored bubbles
$a/R$: (B) 0.22; (C) 0.19; (D) 0.008. The white arrows indicate
missing particle defects at the vertices of the bubble. Scale bars 8
$\mu$$m$.}
\end{figure}

In a typical experiment, the particles adsorbed on a partially
covered bubble are dispersed and exhibit thermal motion (Fig.
\ref{fig1}A). Occasionally, a few particles form transient
aggregates. Analogous equilibrium configurations of colloidal
particles on liquid droplets of fixed volume have been observed
\cite{Bausch03}. In the case of dissolving gas bubbles, the
interparticle distances become smaller until Brownian motion is
arrested, which we term the interfacial jamming transition. This
jammed state can also be reached by packing the bubble surface with
colloidal particles in a microfluidic device \cite{Subramaniam05a}
or by fusing two or more particle-covered bubbles
\cite{Subramaniam05b}. Once the particle movements have stopped, the
bubble does not stabilize but continues to lose gas and deforms away
from a spherical shape (Fig. \ref{fig1}A). It is this nonspherical
bubble that remains stable, as apparently was first observed by
Ramsden \cite{Ramsden}.

We observe that in an air-saturated solution the final non-spherical
shape depends on the ratio of the radius $a$ of the bead and $R$ of
the bubble. For $a/R\sim 0.1$, the bubble exhibits a polyhedral
faceted structure (Fig. \ref{fig1}B-C). The intersection of the
facets is often a missing particle defect, which represents the
position of a five-fold dislocation (white arrows in Fig.
\ref{fig1}B and C). The faceting becomes progressively disordered
until for $a/R\ll 0.1$ the bubble appears highly crumpled (Fig.
\ref{fig1}D).

The measurement of both the local shape of the air-water interface
between the particles and the pressure drop in these micron-size
bubbles remains an experimental challenge. A recent numerical
approach \cite{Lauga04} using Surface Evolver (SE) \cite{Brakke92}
has shown promise in solving the three-dimensional spherical packing
of a small number of particles on emulsion droplets
\cite{Manoharan03}. We perform SE simulations following
\cite{Lauga04} and report here the evolution of the shape for 122
particles on a bubble surface. We have done additional simulations
with almost 400 particles. Solving the full interparticle potential
on the surface is computationally expensive, with simulation time
scaling as the exponential of the number of particles. Here we use
our experimental observations that large-scale rearrangements of the
particles are rare to restrict the interparticle potential
calculation to nearest neighbors and next nearest neighbors, which
makes the simulations with such a large number of particles
tractable.

Particles of volume $V_{p}$ are modeled as liquid droplets embedded
on a larger liquid droplet of volume $V$. The particles have a high
surface tension (typically 30 times larger than the main liquid-gas
surface tension of the bubble) thus maintaining their spherical
shape throughout the simulation. Interfacial tensions of the bubble
and the particles are chosen to satisfy Young's law at the
solid-liquid contact line and to constrain the contact angle to a
fixed value. An exponential repulsive potential is implemented in
order to ensure particle non-interpenetrability (see
\cite{supplement}). To approximate the volume reduction that
accompanies slow dissolution, the volume $V$ of the bubble is
decreased by 2\% increments in each numerical step. SE calculates
the equilibrium configuration of the particles and the shape of the
gas-liquid surface at each step by minimizing the sum of the
gas-liquid surface energies and the total repulsive energy between
the particles.

\begin{figure}
\includegraphics[width=8.5 cm]{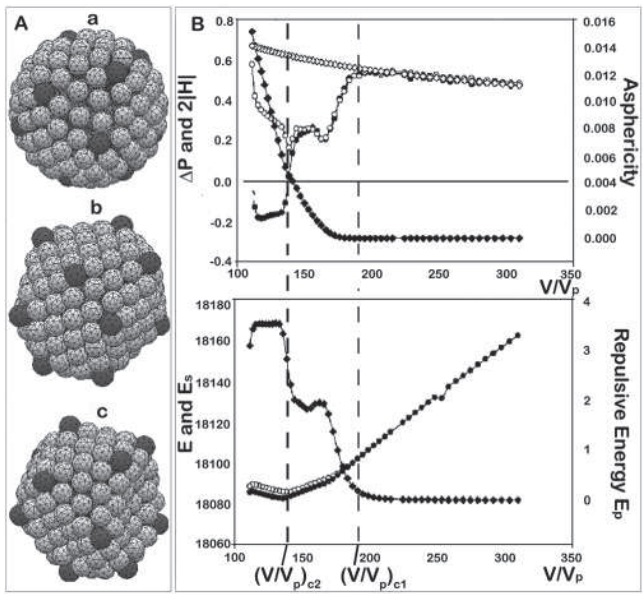}
\caption{\label{fig2} (A)Bubble shapes obtained for $V/V_{p}$ equal
to (a) 203, (b) 138 and (c) 120. The darker particles represent
five-fold dislocation defects. (B) Top graph. Left vertical axis:
($\diamond$) $\Delta P$ without particles, ($\bullet$) $\Delta P$
with particles and ($\circ$) $2\mid H\mid $ versus $V/V_{p}$ where
$H$ is the mean curvature. Right vertical axis: ($\blacklozenge$)
asphericity versus $V/V_{p}$. Bottom graph. Left vertical axis:
($\circ$) total energy E and ($\bullet$) the surface energy $E_{S}$
of the fluid interface versus $V/V_{p}$. Right vertical axis:
($\blacklozenge$) total repulsive energy $E_{P}$ between particles
versus $V/V_{p}$.}
\end{figure}

The simulated armored bubble evolves from a spherical shape (Fig.
\ref{fig2}A(a)) towards a polyhedral shape with facets as $V/V_{p}$
is decreased (Fig. \ref{fig2}A(b)), which matches our experimental
observations. Large volume reductions lead to the inward buckling of
the facets (Fig. \ref{fig2}A(c)). To quantify this observation
further we calculate the asphericity of the bubble \cite{Lidmar03}
which measures the deviation of the shape from that of a perfect
sphere. The asphericity is defined as $\overline {\Delta
R^2}/\overline{R}^2=1/(N\overline{R}^2)\sum_{i=1}^{N}
(R_{i}-\overline{R})^2$, where $N$ is the number of beads, $R_i$ the
distance between the center of the bead $i$ and the center of mass
of all the beads, and $\overline{R}$ the mean radius defined by
$\overline{R}=1/N\sum_{i=1}^N R_{i}$. We observe a sharp increase of
the asphericity when the bubble starts to facet and a significant
change of slope when inward buckling is observed (Fig. \ref{fig2}B).

We next calculate the pressure difference $\Delta P$, obtained
through a native algorithm in SE, between the bubble and its
surroundings as a function of $V/V_{p}$ (Fig. \ref{fig2}B). Unlike a
normal bubble, where $\Delta P$ is a monotonically increasing
function for decreasing $V/V_{p}$ (Fig. \ref{fig2}B), $\Delta P$ of
an armored bubble becomes a decreasing function at $(V/V_{p})_{c1}$,
and eventually reaches zero at $(V/V_{p})_{c2}$. It is significant
that these simulations match the experimentally determined shape of
the pressure curve of millimeter-size particle-covered oil droplets
\cite{Xu05}. Fig. \ref{fig2}B demonstrates the correlation between
$\Delta P$ and the asphericity. Since the particles are held by the
interface, this correlation suggests that the interface is being
deformed as the volume is decreased. Indeed, the absolute value of
the mean curvature $|H|$ of the gas-liquid interface follows exactly
the variation of $\Delta P$ (except near $(V/V_{p})_{c2}$, as SE
gives only $|H|$). It is thus clear that the vanishing of $\Delta P$
is due to the decrease of mean curvature of the gas-liquid interface
towards zero.

Furthermore, we observe that $d\Delta P/dV=dP_{bubble}/dV>0 $ at
$(V/V_{p})_{c1}$, which is a requirement for stability \cite{Kam99}.
For gas-saturated solutions (the case considered in our
experiments), $\Delta P$ also has to go to zero to ensure mechanical
equilibrium. However, more generally chemical potentials must be
equal on either side of the interface \cite{Guggenheim}, which can
be satisfied with $\Delta P\ne 0$. Thus, in the cases of an
oversaturated liquid, the bubble may stabilize at various
intermediate stages of faceting provided that $V/V_p<(V/V_{p})_{c1}$
(the limit being almost no faceting), while in moderately
undersaturated solutions an armored bubble should stabilize with a
highly buckled shape. Thus, in principle, the degree of faceting may
be used as a probe to determine the degree of saturation of the
surrounding medium.

In order to check the stability in terms of energy, we calculate the
total energy $E$, defined as the sum of the total surface energies
$E_{S}$ of all of the interfaces and the total repulsive energy
$E_{P}$ between the particles as a function of $V/V_{p}$ (Fig.
\ref{fig2}B). All energies are normalized by $\gamma L^2$, where $L$
is defined such that $L=V_{p}^{1/3}=(4\pi/3)^{1/3}a$. For comparison
both $E_{S}$ and $E_{P}$ are plotted in Fig. \ref{fig2}B. We observe
that $E$ and $E_{S}$ change slopes as the particles start
interacting, reaching a local minimum at the faceted shape when the
particle interactions are the highest. Inward buckling of the facets
(Fig. \ref{fig2}A(c)) corresponds to a local increase of $E$ in the
energy landscape. Thus, the local minimum is a metastable
equilibrium for this system.

The peculiar ``kink" that the $E_{P}$ curve exhibits during volume
reduction (Fig. \ref{fig2}B) can be traced directly to the packing
of the particles on the surface of the bubble. As the particles are
pushed together during volume reduction, the 12 five-fold defects
serve as the vertices of buckling (dark gray particles in Fig.
\ref{fig2}A) and are pushed away from the center of the bubble. The
increased distance slightly reduces $E_{P}$. This kink in $E_{P}$
also leads to the kink in the $|H|$ and $\Delta P$ curves. We
suggest that the observations on Fig. \ref{fig1} B,C of defects
associated with missing particles of five-fold coordination could
arise from the dewetting and ejection of the particles due to the
higher stresses at these points.

\begin{figure}
\includegraphics[width=8.0 cm]{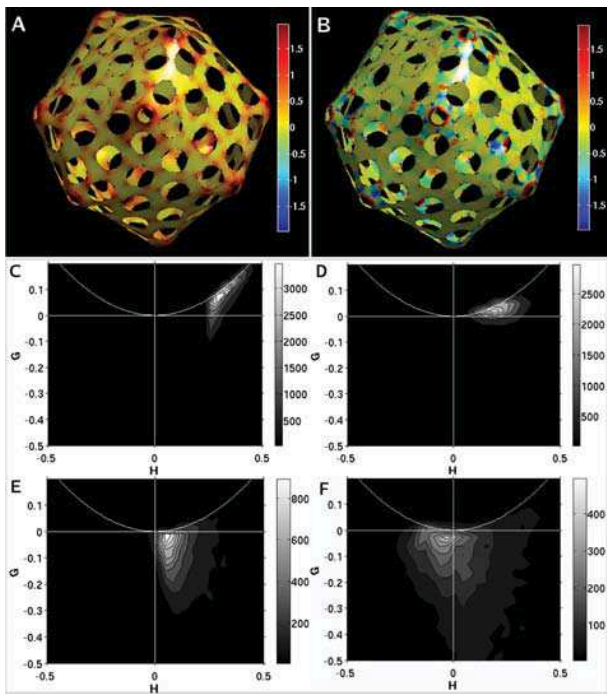}
\caption{\label{fig3} Interfacial distributions of (A) the mean
curvature H and (B) the Gaussian curvature G, for $\rm{V/V_{p}}$
138.1 close to $(V/V_{p})_{c2}$. 2D histogram of the number of
vertices of the interface whose curvatures range between [H,H+0.03]
and [G,G+0.03] obtained for $\rm{V/V_{p}}=$ (C) 253, (D) 162.3, (E)
138.1 and (F) 117.5. Total number of vertices 63016.}
\end{figure}

It appears that the configuration of the gas-liquid interface is
intimately linked to the stability of the armored bubble. We thus
sought to characterize the evolution of the gas-liquid interface
whose shape can be fully specified by the local variation of the
mean curvature $H$ and the Gaussian curvature $G$. Obtaining
accurate local numerical values of $G$ for all simulated $V/V_{p}$
through SE proved impossible at the level of refinement of our
surface due to numerical errors. Thus, we chose four representative
stages in the evolution of the bubble, and systematically refined
the triangulation of the interface to reduce numerical noise. The
spatial distributions of $H$ and $G$ of these surfaces were then
determined with Matlab using algorithms proposed for $H$
\cite{Meyer02} and for $G$ \cite{Goldfeather04}. Representative
images of the interface at approximately $(V/V_{p})_{c2}$ are
reported in a color-coded scheme in Fig. \ref{fig3} A,B. Away from
the particle contact lines the mean curvature $H$ is very nearly
constant, as expected on thermodynamic grounds, and close to zero;
$G$ has a natural distribution since the Gaussian curvature need not
be constant.

Despite the high level of refinement, it is apparent that there is
still some dispersion in $H$ (and in $G$ as well), whose origins are
(i) difficulty in numerical calculations near contact lines and (ii)
errors associated with the triangulation valence around the vertices
which can be amplified during the determination of $G$
\cite{Borrelli03}. Nevertheless, we can draw some conclusions about
the global evolution of the surface curvatures. Indeed, a pair of
values $(H_i, G_i)$ is associated for each vertex defining the
interface. To characterize the global nature of the interface, we
calculate the number density of vertices whose curvatures range
$[H,H+0.03]$ and $[G,G+0.03]$. We report on Fig. \ref{fig3}C-F the
plot of the contour map of this binning of the $H-G$ space; shading
corresponds to the number density of points. As a guide, a sphere
would correspond to a parabola ($G=H^2$) in these plots and the
origin $(0,0)$ corresponds to a planar interface. For large
$V/V_{p}$, when the particles are not interacting, the distribution
of points is concentrated on the parabola, where both $H$ and $G$
are positive (Fig. \ref{fig3}C). Then, for decreasing values of
$V/V_{p}$, the center of the distribution shift towards zero in the
$H$-direction, while in the $G$-direction it becomes negative. These
results indicate a saddle-shape deformation of much of the interface
as the volume is reduced.

We interpret the inward curvature of the interface as a consequence of
Newton's third law. The repulsive beads produce an outward normal
force on each other, because of their confinement on a closed
spherical surface. This outward force must be balanced by an inward
saddle-shaped deformation of the fluid-fluid interface. This
reactive deformation of the interface, which is required for
mechanical equilibrium at each volume reduction, leads to a
reduction in the Laplace pressure. We emphasize
that this saddle-shaped deformation should appear on any initially
spherical fluid-fluid interface carrying repulsive particles, as
soon as the particles are close enough to interact. The details of
the interparticle repulsive force is not relevant for this argument,
the limiting case being the case of a hard sphere repulsion between
the particles, where the interface will deviate from its spherical
shape only when the particles enter into contact.

In conclusion, we have shown that armored bubbles stabilize in
faceted or crumpled shapes by jamming the particles on their
interfaces. Through simulations we demonstrated that the faceted
state is a minimum energy configuration characterized by a mostly
saddle-shaped gas-liquid interface with zero mean curvature. This
minimum is also marked by the vanishing of the Laplace overpressure
$\Delta P$, and $d\Delta P/dV>0$ which guarantees stability against
dissolution. The results we obtained in this study should also be
applicable to describe the interface and behavior of liquid-liquid
systems.

We thank the Harvard MRSEC (DMR-0213805) and Unilever Research for
support and D. Gregory for helpful conversations. SHK and SMY were
supported by the Creative Research Initiative Program of MOST/KOSEF
and the BK21 program. We thank E. Lauga for help with SE.
\bibliography{Prlbiblio}

\begin{thebibliography}{28}
\expandafter\ifx\csname natexlab\endcsname\relax\def\natexlab#1{#1}\fi
\expandafter\ifx\csname bibnamefont\endcsname\relax
  \def\bibnamefont#1{#1}\fi
\expandafter\ifx\csname bibfnamefont\endcsname\relax
  \def\bibfnamefont#1{#1}\fi
\expandafter\ifx\csname citenamefont\endcsname\relax
  \def\citenamefont#1{#1}\fi
\expandafter\ifx\csname url\endcsname\relax
  \def\url#1{\texttt{#1}}\fi
\expandafter\ifx\csname urlprefix\endcsname\relax\def\urlprefix{URL }\fi
\providecommand{\bibinfo}[2]{#2}
\providecommand{\eprint}[2][]{\url{#2}}

\bibitem[{\citenamefont{W.Ramsden}(1903)}]{Ramsden}
\bibinfo{author}{\bibnamefont{W.Ramsden}}, \bibinfo{journal}{Proc. R. Soc.
  London} \textbf{\bibinfo{volume}{72}}, \bibinfo{pages}{156}
  (\bibinfo{year}{1903}).

\bibitem[{\citenamefont{Johnson and Cooke}(1981)}]{Johnson81}
\bibinfo{author}{\bibfnamefont{B.}~\bibnamefont{Johnson}} \bibnamefont{and}
  \bibinfo{author}{\bibfnamefont{R.}~\bibnamefont{Cooke}},
  \bibinfo{journal}{Science} \textbf{\bibinfo{volume}{213}},
  \bibinfo{pages}{209} (\bibinfo{year}{1981}).

\bibitem[{\citenamefont{Du et~al.}(2003)\citenamefont{Du, Bilbao-Montoya,
  Binks, Dickinson, Ettelaie, and Murray}}]{Du03}
\bibinfo{author}{\bibfnamefont{Z.}~\bibnamefont{Du}},
  \bibinfo{author}{\bibfnamefont{M.}~\bibnamefont{Bilbao-Montoya}},
  \bibinfo{author}{\bibfnamefont{B.}~\bibnamefont{Binks}},
  \bibinfo{author}{\bibfnamefont{E.}~\bibnamefont{Dickinson}},
  \bibinfo{author}{\bibfnamefont{R.}~\bibnamefont{Ettelaie}}, \bibnamefont{and}
  \bibinfo{author}{\bibfnamefont{B.}~\bibnamefont{Murray}},
  \bibinfo{journal}{Langmuir} \textbf{\bibinfo{volume}{19}},
  \bibinfo{pages}{3106} (\bibinfo{year}{2003}).

\bibitem[{\citenamefont{Dickinson et~al.}(2004)\citenamefont{Dickinson,
  Ettelaie, Kostakis, and Murray}}]{Dickinson04}
\bibinfo{author}{\bibfnamefont{E.}~\bibnamefont{Dickinson}},
  \bibinfo{author}{\bibfnamefont{R.}~\bibnamefont{Ettelaie}},
  \bibinfo{author}{\bibfnamefont{T.}~\bibnamefont{Kostakis}}, \bibnamefont{and}
  \bibinfo{author}{\bibfnamefont{B.}~\bibnamefont{Murray}},
  \bibinfo{journal}{Langmuir} \textbf{\bibinfo{volume}{20}},
  \bibinfo{pages}{8517} (\bibinfo{year}{2004}).

\bibitem[{\citenamefont{Alargova et~al.}(2004)\citenamefont{Alargova,
  Warhadpande, Paunov, and Velev}}]{Alargova04}
\bibinfo{author}{\bibfnamefont{R.}~\bibnamefont{Alargova}},
  \bibinfo{author}{\bibfnamefont{D.}~\bibnamefont{Warhadpande}},
  \bibinfo{author}{\bibfnamefont{V.}~\bibnamefont{Paunov}}, \bibnamefont{and}
  \bibinfo{author}{\bibfnamefont{O.}~\bibnamefont{Velev}},
  \bibinfo{journal}{Langmuir} \textbf{\bibinfo{volume}{20}},
  \bibinfo{pages}{10371} (\bibinfo{year}{2004}).

\bibitem[{\citenamefont{Binks and Horozov}(2005)}]{Binks05}
\bibinfo{author}{\bibfnamefont{B.}~\bibnamefont{Binks}} \bibnamefont{and}
  \bibinfo{author}{\bibfnamefont{T.}~\bibnamefont{Horozov}},
  \bibinfo{journal}{Angew. Chem.} \textbf{\bibinfo{volume}{44}},
  \bibinfo{pages}{3722} (\bibinfo{year}{2005}).

\bibitem[{\citenamefont{Schutt et~al.}(2003)\citenamefont{Schutt, Klein,
  Mattrey, and Reiss}}]{Schutt05}
\bibinfo{author}{\bibfnamefont{E.}~\bibnamefont{Schutt}},
  \bibinfo{author}{\bibfnamefont{D.}~\bibnamefont{Klein}},
  \bibinfo{author}{\bibfnamefont{R.}~\bibnamefont{Mattrey}}, \bibnamefont{and}
  \bibinfo{author}{\bibfnamefont{J.}~\bibnamefont{Reiss}},
  \bibinfo{journal}{Angew. Chem.} \textbf{\bibinfo{volume}{42}},
  \bibinfo{pages}{3218} (\bibinfo{year}{2003}).

\bibitem[{\citenamefont{Krachelsvky and Nagayama}(2001)}]{Krachelsky01}
\bibinfo{author}{\bibfnamefont{P.}~\bibnamefont{Krachelsvky}} \bibnamefont{and}
  \bibinfo{author}{\bibfnamefont{K.}~\bibnamefont{Nagayama}}
  (\bibinfo{publisher}{Elsevier Science}, \bibinfo{address}{New York},
  \bibinfo{year}{2001}).

\bibitem[{\citenamefont{Adamson and Gast}(1997)}]{Adamson97}
\bibinfo{author}{\bibfnamefont{A.}~\bibnamefont{Adamson}} \bibnamefont{and}
  \bibinfo{author}{\bibfnamefont{A.}~\bibnamefont{Gast}}
  (\bibinfo{publisher}{Wiley-Interscience}, \bibinfo{address}{New York},
  \bibinfo{year}{1997}), \bibinfo{edition}{6th} ed.

\bibitem[{\citenamefont{Gibbs et~al.}(1999)\citenamefont{Gibbs, Alli, and
  Mulligan}}]{Gibbs99}
\bibinfo{author}{\bibfnamefont{B.}~\bibnamefont{Gibbs}},
  \bibinfo{author}{\bibfnamefont{S.~K.~I.} \bibnamefont{Alli}},
  \bibnamefont{and} \bibinfo{author}{\bibfnamefont{C.}~\bibnamefont{Mulligan}},
  \bibinfo{journal}{Int. J. Food Sci. Nutr.} \textbf{\bibinfo{volume}{50}},
  \bibinfo{pages}{213} (\bibinfo{year}{1999}).

\bibitem[{\citenamefont{Duncan and Needham}(2004)}]{Needham}
\bibinfo{author}{\bibfnamefont{P.~B.} \bibnamefont{Duncan}} \bibnamefont{and}
  \bibinfo{author}{\bibfnamefont{D.}~\bibnamefont{Needham}},
  \bibinfo{journal}{Langmuir} \textbf{\bibinfo{volume}{20}},
  \bibinfo{pages}{2567} (\bibinfo{year}{2004}).

\bibitem[{\citenamefont{Hilden and Trumble}(2003)}]{Hilden03}
\bibinfo{author}{\bibfnamefont{J.}~\bibnamefont{Hilden}} \bibnamefont{and}
  \bibinfo{author}{\bibfnamefont{K.}~\bibnamefont{Trumble}},
  \bibinfo{journal}{J. Colloid Interface Sci.} \textbf{\bibinfo{volume}{267}},
  \bibinfo{pages}{463} (\bibinfo{year}{2003}).

\bibitem[{\citenamefont{Kam and Rossen}(1999)}]{Kam99}
\bibinfo{author}{\bibfnamefont{S.}~\bibnamefont{Kam}} \bibnamefont{and}
  \bibinfo{author}{\bibfnamefont{W.}~\bibnamefont{Rossen}},
  \bibinfo{journal}{J. Colloid Interface Sci.} \textbf{\bibinfo{volume}{213}},
  \bibinfo{pages}{329} (\bibinfo{year}{1999}).

\bibitem[{\citenamefont{Subramanian et~al.}(2005)\citenamefont{Subramanian,
  Larsen, and Stone}}]{Subramanian05}
\bibinfo{author}{\bibfnamefont{R.}~\bibnamefont{Subramanian}},
  \bibinfo{author}{\bibfnamefont{R.}~\bibnamefont{Larsen}}, \bibnamefont{and}
  \bibinfo{author}{\bibfnamefont{H.}~\bibnamefont{Stone}},
  \bibinfo{journal}{Langmuir} \textbf{\bibinfo{volume}{21}},
  \bibinfo{pages}{4526} (\bibinfo{year}{2005}).

\bibitem[{\citenamefont{Subramaniam et~al.}(2006)\citenamefont{Subramaniam,
  M\'ejean, Abkarian, and Stone}}]{Subramaniam06}
\bibinfo{author}{\bibfnamefont{A.~B.} \bibnamefont{Subramaniam}},
  \bibinfo{author}{\bibfnamefont{C.}~\bibnamefont{M\'ejean}},
  \bibinfo{author}{\bibfnamefont{M.}~\bibnamefont{Abkarian}}, \bibnamefont{and}
  \bibinfo{author}{\bibfnamefont{H.}~\bibnamefont{Stone}},
  \bibinfo{journal}{Langmuir} \textbf{\bibinfo{volume}{22}},
  \bibinfo{pages}{5986} (\bibinfo{year}{2006}).

\bibitem[{sup()}]{supplement}
\bibinfo{note}{EPAPS supplementary information.}

\bibitem[{\citenamefont{Bausch et~al.}(2003)\citenamefont{Bausch, Bowick,
  Cacciuto, Dinsmore, Hsu, Nelson, Nikolaides, Travesset, and
  Weitz}}]{Bausch03}
\bibinfo{author}{\bibfnamefont{A.}~\bibnamefont{Bausch}},
  \bibinfo{author}{\bibfnamefont{M.}~\bibnamefont{Bowick}},
  \bibinfo{author}{\bibfnamefont{A.}~\bibnamefont{Cacciuto}},
  \bibinfo{author}{\bibfnamefont{A.}~\bibnamefont{Dinsmore}},
  \bibinfo{author}{\bibfnamefont{M.}~\bibnamefont{Hsu}},
  \bibinfo{author}{\bibfnamefont{D.}~\bibnamefont{Nelson}},
  \bibinfo{author}{\bibfnamefont{M.}~\bibnamefont{Nikolaides}},
  \bibinfo{author}{\bibfnamefont{A.}~\bibnamefont{Travesset}},
  \bibnamefont{and} \bibinfo{author}{\bibfnamefont{D.}~\bibnamefont{Weitz}},
  \bibinfo{journal}{Science} \textbf{\bibinfo{volume}{299}},
  \bibinfo{pages}{1716} (\bibinfo{year}{2003}).

\bibitem[{\citenamefont{Subramaniam
  et~al.}(2005{\natexlab{a}})\citenamefont{Subramaniam, Abkarian, and
  Stone}}]{Subramaniam05a}
\bibinfo{author}{\bibfnamefont{A.~B.} \bibnamefont{Subramaniam}},
  \bibinfo{author}{\bibfnamefont{M.}~\bibnamefont{Abkarian}}, \bibnamefont{and}
  \bibinfo{author}{\bibfnamefont{H.}~\bibnamefont{Stone}},
  \bibinfo{journal}{Nat. Mat.} \textbf{\bibinfo{volume}{4}},
  \bibinfo{pages}{553} (\bibinfo{year}{2005}{\natexlab{a}}).

\bibitem[{\citenamefont{Subramaniam
  et~al.}(2005{\natexlab{b}})\citenamefont{Subramaniam, Abkarian, Mahadevan,
  and Stone}}]{Subramaniam05b}
\bibinfo{author}{\bibfnamefont{A.~B.} \bibnamefont{Subramaniam}},
  \bibinfo{author}{\bibfnamefont{M.}~\bibnamefont{Abkarian}},
  \bibinfo{author}{\bibfnamefont{L.}~\bibnamefont{Mahadevan}},
  \bibnamefont{and} \bibinfo{author}{\bibfnamefont{H.}~\bibnamefont{Stone}},
  \bibinfo{journal}{Nature} \textbf{\bibinfo{volume}{438}},
  \bibinfo{pages}{930} (\bibinfo{year}{2005}{\natexlab{b}}).

\bibitem[{\citenamefont{Lauga and Brenner}(2004)}]{Lauga04}
\bibinfo{author}{\bibfnamefont{E.}~\bibnamefont{Lauga}} \bibnamefont{and}
  \bibinfo{author}{\bibfnamefont{M.}~\bibnamefont{Brenner}},
  \bibinfo{journal}{Phys. Rev. Lett.} \textbf{\bibinfo{volume}{93}},
  \bibinfo{pages}{238301} (\bibinfo{year}{2004}).

\bibitem[{\citenamefont{Brakke}(1992)}]{Brakke92}
\bibinfo{author}{\bibfnamefont{K.}~\bibnamefont{Brakke}},
  \bibinfo{journal}{Exp. Math.} \textbf{\bibinfo{volume}{1}},
  \bibinfo{pages}{141} (\bibinfo{year}{1992}).

\bibitem[{\citenamefont{Manoharan et~al.}(2003)\citenamefont{Manoharan,
  Elsesser, and Pine}}]{Manoharan03}
\bibinfo{author}{\bibfnamefont{V.}~\bibnamefont{Manoharan}},
  \bibinfo{author}{\bibfnamefont{M.}~\bibnamefont{Elsesser}}, \bibnamefont{and}
  \bibinfo{author}{\bibfnamefont{D.}~\bibnamefont{Pine}},
  \bibinfo{journal}{Science} \textbf{\bibinfo{volume}{301}},
  \bibinfo{pages}{483} (\bibinfo{year}{2003}).

\bibitem[{\citenamefont{Lidmar et~al.}(2003)\citenamefont{Lidmar, Mirny, and
  Nelson}}]{Lidmar03}
\bibinfo{author}{\bibfnamefont{J.}~\bibnamefont{Lidmar}},
  \bibinfo{author}{\bibfnamefont{L.}~\bibnamefont{Mirny}}, \bibnamefont{and}
  \bibinfo{author}{\bibfnamefont{D.}~\bibnamefont{Nelson}},
  \bibinfo{journal}{Phys. Rev. E} \textbf{\bibinfo{volume}{68}},
  \bibinfo{pages}{051910} (\bibinfo{year}{2003}).

\bibitem[{\citenamefont{Xu et~al.}(2005)\citenamefont{Xu, Melle, Golemanov, and
  Fuller}}]{Xu05}
\bibinfo{author}{\bibfnamefont{H.}~\bibnamefont{Xu}},
  \bibinfo{author}{\bibfnamefont{S.}~\bibnamefont{Melle}},
  \bibinfo{author}{\bibfnamefont{K.}~\bibnamefont{Golemanov}},
  \bibnamefont{and} \bibinfo{author}{\bibfnamefont{G.}~\bibnamefont{Fuller}},
  \bibinfo{journal}{Langmuir} \textbf{\bibinfo{volume}{21}},
  \bibinfo{pages}{10016} (\bibinfo{year}{2005}).

\bibitem[{\citenamefont{Guggenheim}(1967)}]{Guggenheim}
\bibinfo{author}{\bibfnamefont{E.}~\bibnamefont{Guggenheim}}
  (\bibinfo{publisher}{North-Holland, Elsevier}, \bibinfo{address}{Amsterdam},
  \bibinfo{year}{1967}), \bibinfo{edition}{4th} ed.

\bibitem[{\citenamefont{Meyer et~al.}(2002)\citenamefont{Meyer, Desbrun,
  Schroder, and Barr}}]{Meyer02}
\bibinfo{author}{\bibfnamefont{M.}~\bibnamefont{Meyer}},
  \bibinfo{author}{\bibfnamefont{M.}~\bibnamefont{Desbrun}},
  \bibinfo{author}{\bibfnamefont{P.}~\bibnamefont{Schroder}}, \bibnamefont{and}
  \bibinfo{author}{\bibfnamefont{A.}~\bibnamefont{Barr}},
  \emph{\bibinfo{title}{{\rm in} International workshop on visualization and
  mathematics}} (\bibinfo{publisher}{Berlin-Dahlem, Germany},
  \bibinfo{year}{2002}).

\bibitem[{\citenamefont{Goldfeather and Interrante}(2004)}]{Goldfeather04}
\bibinfo{author}{\bibfnamefont{J.}~\bibnamefont{Goldfeather}} \bibnamefont{and}
  \bibinfo{author}{\bibfnamefont{V.}~\bibnamefont{Interrante}},
  \bibinfo{journal}{ACM Trans. Graph.} \textbf{\bibinfo{volume}{23}},
  \bibinfo{pages}{45} (\bibinfo{year}{2004}).

\bibitem[{\citenamefont{Borrelli et~al.}(2003)\citenamefont{Borrelli, Cazals,
  and Morvan}}]{Borrelli03}
\bibinfo{author}{\bibfnamefont{V.}~\bibnamefont{Borrelli}},
  \bibinfo{author}{\bibfnamefont{F.}~\bibnamefont{Cazals}}, \bibnamefont{and}
  \bibinfo{author}{\bibfnamefont{J.-M.} \bibnamefont{Morvan}},
  \bibinfo{journal}{Comput. Aided Geom. Design} \textbf{\bibinfo{volume}{20}},
  \bibinfo{pages}{1} (\bibinfo{year}{2003}).

\end{thebibliography}

\end{document}